\documentclass[useAMS,usenatbib]{mn2e}
\usepackage{graphicx}

\usepackage{subfigure}
\usepackage{amsmath}
\usepackage{mathabx}

\title[The Baryon Cycle of Disc Galaxies]{MaGICC Baryon Cycle: 
The Enrichment History of Simulated Disc Galaxies}
\author[C.B. Brook et al.]{C.~B. Brook$^{1}$, G.~Stinson$^{2}$, 
B.~K. Gibson$^{3,4}$, S.~Shen$^{5}$,  A.~V.~Macci\`{o}$^{2}$,\newauthor 
 A. Obreja$^{1}$, J.~Wadsley$^{6}$, T. Quinn$^{7}$\\
$^1$Departamento de F\'{i}sica Te\'{o}rica, Universidad Aut\'{o}noma de 
Madrid, E-28049 Cantoblanco, Madrid, Spain\\
$^2$Max-Planck-Institut f\"ur Astronomie, K\"onigstuhl 17, 69117 
Heidelberg, Germany\\
$^3$Jeremiah Horrocks Institute, University of Central Lancashire, 
Preston, PR1~2HE, UK \\
$^4$Department of Astronomy \& Astrophysics, University of California, 
1156 High Street, Santa Cruz, CA 95064, USA\\
$^5$Institute for Computational Astrophysics and Dept of Astronomy \& Astrophysics, Saint MaryÕs University, Halifax, NS, B3H 3C3, Canada\\
$^6$Department of Physics \& Astronomy, McMaster University, Hamilton, 
Ontario, L8S~4M1, Canada\\
$^7$Astronomy Department, University of Washington, 
Box 351580, Seattle, WA 98195-1580, USA} 

\begin{document}
\date{Submitted}
\pagerange{\pageref{firstpage}--\pageref{lastpage}} 
\pubyear{2013}

\maketitle
\label{firstpage}

\begin{abstract}
Using cosmological galaxy formation simulations from the MaGICC 
project, spanning stellar mass from $\sim$10$^7$\,M$_\odot$ to 3$\times$10$^{10}$\,M$_\odot$, we trace the 
baryonic cycle of infalling gas from the virial radius through to its 
eventual participation in the star formation process.  An emphasis is 
placed upon the temporal history of chemical enrichment during its 
passage through the corona and circum-galactic medium. We derive the 
distributions of time between gas crossing the virial radius and being 
accreted to the star forming region (which allows for mixing within the 
corona), as well as the time between gas being accreted to the star 
forming region and then ultimately forming stars (which allows for 
mixing within the disc).  Significant numbers of stars are 
formed from gas that cycles back through the hot halo after first 
accreting to the star forming region. Gas entering high mass galaxies is 
pre-enriched in low mass proto-galaxies prior to entering the virial 
radius of the central progenitor, with only small amounts of primordial 
gas accreted, even at high redshift ($z$$\sim$5). After entering the 
virial radius, significant further enrichment occurs prior to the 
accretion of the gas to the star forming region, with gas that is 
feeding the star forming region surpassing 0.1\,Z$_\odot$ by $z$=0. 
Mixing with halo gas, itself  enriched via galactic fountains, 
is thus crucial in determining the metallicity at which gas is accreted 
to the disc.  The lowest mass simulated galaxy 
(M$_{vir}$$\sim$$2$$\times$10$^{10}$\,M$_\odot$, with 
M$_{*}$$\sim$10$^{7}$\,M$_\odot$), by contrast, accretes primordial gas 
through the virial radius and onto the disc, throughout its history. 
Much like the case for classical analytical solutions to the so-called 
`G-dwarf problem', overproduction of low-metallicity stars is 
ameliorated by the interplay between the time of accretion onto the disc 
and the subsequent involvement in star formation - i.e., due to the 
inefficiency of star formation.  Finally, gas outflow / metal removal 
rates from star forming regions as a function of galactic mass are 
presented.
\end{abstract}

\begin{keywords}
galaxies: evolution -- galaxies: formation -- galaxies: spiral
\end{keywords}

\section{Introduction}

\begin{table*}
\label{tab:data}
\begin{minipage}{180mm}
\begin{center}
\caption{Simulation data where gas part is the initial gas particle mass, $c_\star$ is the star formation efficiency parameter, C$_{\rm Diff}$ is the diffusion co-efficient, M$_*$ and M$_{halo}$ are measured within the virial radius,   $M_R$ is the R band magnitude calculated using SUNRISE,  R$_{\rm vir}$ is at  390 times the cosmic background matter density, {\it acc.} is the fraction of the universal baryon fraction that accrete to R$_{\rm vir}$ during the simulation,  while $z=0$ shows the fraction remaining within  R$_{\rm vir}$ at the final timestep, {\it prim. frac} is the fraction of gas that accretes to  R$_{\rm vir}$  that is primordial, and  {\it fount. frac.} is the fraction of  stars forming from gas that was cycled through the galactic fountain, requiring that the gas is expelled beyond  R$_{\rm vir}$/8 and subsequently re-accreted to the star forming region.  }
\begin{tabular}{llllllllllrllll}
\hline
Name & MUGS  &gas part. &$c_\star$ &C$_{\rm Diff}$&M$_{halo}$ &R$_{\rm vir}$ &M$_\star$  & M$_{gas}$ &$M_R$&\multicolumn{2}{c}{baryon frac.}&R$_{\rm vir/8}$ &prim.&  fount.\\
         &label & [M$_\odot$]  &   &  & [M$_\odot$]&[kpc] &[M$_\odot$]& [M$_\odot$]& &acc.& $z$$=$$0$&acc. &frac. & frac.\\
\hline
dG3  & 15784  &   3.1$\times$10$^3$   &0.017 &0.05 & $2.2$$\times$10$^{10}$ &77 &$8.9$$\times$10$^{6}$ &  $4.4$$\times$10$^{8}$ &-12.8 &0.46& 0.15& 0.14& 0.95& 0.22\\
SG1   & 5664    &  2.5$\times$10$^4$   &0.017& 0.05 & $6.5$$\times$10$^{10}$ & 115&$2.3$$\times$10$^{8}$ &  $4.7$$\times$10$^{9}$ &-17.0 & 0.85&0.50& 0.55&0.62&0.37\\
SG2   & 1536    &  2.5$\times$10$^4$    & 0.017 & 0.05&$8.3$$\times$10$^{10}$ &125  &$4.5$$\times$10$^{8}$ &  $6.2$$\times$10$^{9}$ &-17.5 & 1.0&0.52& 0.66&0.57&0.39\\
SG3    & 15784 &  2.5$\times$10$^4$    &0.017 & 0.05 & $1.8$$\times$10$^{11}$ & 162&$4.2$$\times$10$^{9}$ &  $1.5$$\times$10$^{10}$ &-20.0 &1.0  &0.73&0.78&0.53&0.32\\
SG3$_{\rm LD}$  & 15784  & 2.5$\times$10$^4$ &0.017 &0.01 & $1.8$$\times$10$^{11}$&163 &  $3.7$$\times$10$^{9}$ & $1.7$$\times$10$^{10}$ &-19.8   &1.0&0.76&0.77&0.57&0.32\\
SG4   & 15807 &  2.5$\times$10$^4$  &0.017&0.05  &$3.2$$\times$10$^{11}$ &190 &$1.4$$\times$10$^{10}$ & $2.8$$\times$10$^{10}$ & -21.2 &1.0 &0.85&0.80& 0.50&0.32\\
L$^*$G2    & 1536      &   2.0$\times$10$^5$   &0.033  &0.05 &$7.6$$\times$10$^{11}$ &257 &$3.0$$\times$10$^{10}$ &  $6.6$$\times$10$^{10}$ &-21.7&1.0&0.82&0.69&0.08& 0.18\\
L$^*$G2$_{\rm LD}$    & 1536      &   2.0$\times$10$^5$  &0.033 &0.01  &$7.6$$\times$10$^{11}$&258 & $2.7$$\times$10$^{10}$ &  $7.1$$\times$10$^{10}$ &-21.6  &1.0&0.87&0.69&0.22&0.18\\
\hline
\end{tabular}\\
\end{center}
\end{minipage}
\end{table*}

The formation, accretion, and expulsion of metals leaves important 
chemical imprints within the stellar and gas-phase components of 
galaxies. These signatures, including, for example, abundance ratios, 
metallicity distribution functions, and spatially-resolved abundance 
gradients, can be used to constrain the  baryonic cycle within 
galaxies - that is, the infall of gas into a system, its involvement in 
star formation, and its potential outflow via energy- and/or 
momentum-driven winds.  This cycle, and how it is affected by the mass, 
environment, and accretion history of the host halo in which a galaxy 
resides, lies at the heart of galaxy formation.

Chemical evolution models follow the creation and evolution of metals, 
accounting for the accretion of pristine and pre-enriched gas at rates 
that are constrained in order to match the observed chemical abundance 
properties of galaxies 
\citep[e.g.][]{timmes95,gibson97,chiappini01,fenner03}, as well as 
properties such as luminosities, star formation rates, and colours. 
Whilst providing important insights, such models lack the 
dynamical/kinematic information required for a more comprehensive model 
of galaxy formation. Models such as \cite{samland97} incorporate chemical evolution into a dynamical model of the galaxy, including the mixing of metals in the ISM, and in the case of \cite{samland03}, within a growing dark matter halo.  One way that chemical evolution modelling can 
incorporate the full merging and mass evolution history within a 
cosmological context is via Semi-Analytic Models (SAMs).  Gas accretion 
and, therefore, star formation, are tied to the growth and merging 
histories of dark matter halos, thereby allowing chemical enrichment to 
be traced \citep[e.g.][]{calura09,arrigoni10}. See \cite{benson10} for a  review of various ways of modelling galaxy formation,  and their advantages and disadvantages. 

An even more self-consistent framework can be constructed by embedding 
star formation, feedback, and chemical enrichment within fully 
cosmological hydrodynamical simulations of galaxy formation. Indeed, 
there is a rich vein of literature linking chemical evolution and 
hydrodynamics in such a manner 
\citep[e.g.][]{steinmetz94,berczik99,springel03,kawata03,bailin05,renda05,martinez08,sawala10,tissera12,kobayashi11,few12}.

However, simulations on cosmological scales can not resolve the complex processes occurring within a multi- phase inter stellar medium (ISM), where the volume is dominated by hot, diffuse gas while most of the mass lies in cold, dense clouds.  Several methods of modelling the ISM within the sub-grid have been developed. \cite{hultman99} used hot and cold phases,  and allowed them to interact through radiative cooling and evaporation of the cold clouds. \cite{pearce01}  set temperature boundaries to  decouple cold and hot phases.  \cite{springel03} developed an analytic model for the sub-grid scale,  regulating star formation in a multiphase ISM within a simulation particle. \cite{semelin02}  consider a warm gas phase, treated as a continuous fluid using SPH, and a cold gas phase, treated by a low-dissipation sticky particle component.  \cite{harfst06} also used a multi-phase ISM, including condensation, evaporation, drag, and energy dissipation and a  star formation efficiency that is dependent on the ISM properties. \cite[][see also \citealt{marri03}]{scannapieco06} include a scheme whereby particles with different thermodynamic  properties do not see each other as neighbours,  allowing hot, diffuse gas to coexist with cold, dense gas.
\cite{pelupessy06} formulate a subgrid model for gas clouds that use cloud scaling relations, and tracks the formation of H2 on dust grains and its destruction by UV irradiation, including the shielding by dust and H2 self-shielding, as well as its collisional destruction in the warm neutral medium. 
\cite{thacker01} disallowed radiative losses for 30\,Myr from particles which had just been heated by SNe,   with hot and cold gas co-existing on larger scales, and  allowing a wind to develop,. \cite{stinson06} developed this model further by using the blast-wave model for supernova \citep{mckee77} to  relate the timescale of the `adiabatic' phase to the local density and pressure of the star forming region.

 Within our Making Galaxies in a Cosmological Context (MaGICC) project we have simulated a suite of galaxies using the 
cosmological hydro-dynamical galaxy formation code \textsc{Gasoline},  and included blast-wave  supernova feedback,  as well as ionising feedback from massive stars prior to their explosion as supernovae. As discussed in our earlier work, these MaGICC galaxies have the star formation and feedback parameters (see Section~\ref{code})
tuned to match the stellar mass-halo mass relation.
The simulations then
match a wide range of galaxy scaling relations relations \citep{brook12b},  over a stellar mass 
ranging from 2.3$\times$10$^{8}$\,M$_\odot$ to 3$\times$10$^{10}~$M$_\odot$
(Table~1).  Further, the simulations 
expel sufficient metals to match local observations 
\citep{prochaska11,tumlinson11} of OVI in the circum-galactic medium 
\citep{stinson12,brook12c}, and have been shown to match the {\it 
evolution} of the stellar mass-halo mass relation 
\citep{stinson13,kannan13}, as derived in abundance matching studies 
\citep{moster13}. 

We showed in \citet{pilkington12b}, 
and return to below, our MaGICC simulations do not suffer from any 
`G-dwarf problem' - i.e., an overproduction of low-metallicity stars 
relative to the number observed in nature. Further constraints on our simulations are provided by so called `near-field cosmology': our simulations have appropriately low mass stellar halos for their mass and morphological type \cite[see][]{brook04a}, and abundances ratios which mimic those found in the Milky Way thick and thin discs \cite{brook12c, stinson13b}

This confluence of 
simulations with observations places necessary constraints on the particular  baryon cycle,  a cycle which can be measured directly within simulations. What we wonder is, for models that match a range of z=0 relations \cite[see also][which match at least a subset of the relations matched by our simulations]{mccarthy12,aumer13}, how many degeneracies in the baryon cycle are possible? If a fundamentally different feedback implementations also matches z=0 observations, can it have totally different baryon cycle to ours? By providing quantitative measures of this cycle, we aim to facilitate comparisons with other models, particularly those employing significantly different implementations of  star formation and feedback processes, as well as providing  predictions for observers.

Previously, \cite{shen10} showed that in a Milky Way-analogue 
simulation, intergalactic medium metals primarily reside in the 
so-called warm-hot intergalactic medium (WHIM) with metallicities lying 
between 0.01 and 0.1 solar with a slight decrease at lower redshifts. In 
galaxies of such mass, enrichment of the WHIM by proto-galaxies at high 
redshift means that the majority of gas is pre-enriched prior to 
accretion to the central galaxy.

Here, we track the baryon cycle of MaGICC galaxies which span more than 
three orders-of-magnitude in stellar mass, and present the inflow and 
outflow rates of their gas and metals. We show the importance of (a) 
pre-enrichment, and give an indication of the mass range at which it 
becomes important, and link it to the existence of an accreted stellar 
halo, (b) enrichment through gas mixing within the halo itself, both 
during the first accretion to the disc and during galactic fountain 
cycles, and (c) enrichment within the disc region itself. First, we 
present the details of the code \textsc{Gasoline} and the properties of 
the simulated galaxaes in Section~\ref{code}. In Section~\ref{inflows}, 
we explore the evolution of the inflow rates through the virial radius 
R$_{\rm{vir}}$, and into the star forming regions (through 
R$_{\rm{vir}}/8$\footnote{This value is somewhat arbitrary, but is generally a reasonable indication of the region where star formation occurs.}), and the metallicities of such inflowing gas. In 
Sections~\ref{timescales}\,\&\,\ref{recycle}, we provide the timescales 
of gas entering the virial radius, entering the star forming 
region, cycling though the galactic fountain, and finally forming stars. 
In Section~\ref{outflows}, we examine the evolution of the rates of 
outflow from the star forming regions (through R$_{\rm{vir}}/8$) and 
their metallicities. We end by discussing our results and their 
implications in Section~\ref{summary}.

\section{The Simulations}
\label{code}

The MaGICC simulations were realised using \text{Gasoline} 
\citep{wadsley04}, a fully parallel, gravitational N-body + smoothed 
particle hydrodynamics (SPH) code.  Cooling via hydrogen, helium, and 
various metal-lines is included, after \citet{shen10}, employing 
\textsc{CLOUDY} (v.07.02 \citealt{ferland98}), assuming ionisation 
equilibrium and cooling rates self-consistently, in the presence of a 
uniform ultraviolet ionising background \citep{haardtmadau96}. 

We do not follow the atomic-to-molecular
transition. Simulations of dwarf galaxies with H2-
regulated star formation and comparable resolution to our low mass galaxies have been run to z = 0 by \cite{christensen12}. However, the resolution of our most massive galaxies precludes the inclusion of such processes. 
We prevent gas from collapsing to higher densities than SPH can physically 
resolve: (i) pressure is added to the gas in high density star forming 
regions \citep{robertson08}, to ensure that gas resolves the Jeans mass 
and does not artificially fragment, and (ii) a maximum density limit is 
imposed by setting a minimum SPH smoothing length of 0.25 times that of 
the gravitational softening length.

The simulations described here are cosmological zoom simulations derived 
from the McMaster Unbiased Galaxy Simulations (MUGS: 
\citealt{stinson10}),  which use WMAP 3 cosmology \cite{spergel07} and follow evolution from $z$$=$99 to $z$$=$0.  For SG1 to SG4 (see Table~1), the initial 
conditions are `scaled down' variants of those employed in the original 
MUGS work, so that rather than residing in a 68\,Mpc cube, they lie 
within a cube with 34\,Mpc sides (and for dG3, they are scaled down 
further to a cube with 17\,Mpc sides). This resizing allows us to compare 
galaxies with exactly the same merger histories at a variety of masses. 
Differences in the underlying power spectrum that result from this 
rescaling are minor \citep{springel08,maccio08,kannan12}, and do not 
significantly affect our results. Certainly, if a statistical sample 
were being examined this rescaling may be important, but for the 
purposes of this study, only small quantitive differences would 
eventuate.

\subsection{Star Formation and Feedback}

All simulations use the feedback scheme as described in 
\citet{stinson12,stinson13}, which we describe briefly here.  Gas is 
eligible to form stars when it reaches cool temperatures ($T$$<$$15,000$ 
K) in a dense environment ($n_{th}$$>$$9.3$ cm$^{-3}$); the latter is set 
to be the maximum density that gas can reach using gravity - i.e., 32 
m$_{gas}/\epsilon{^3}$. We note that in all our simulations, stars form at temperatures significantly below the threshold temperature, meaning  that it is the threshold density that is critical in determining whether gas is eligible for  star formation.  Such gas is converted to stars according to a 
Schmidt Law:
\begin{equation}
\frac{\Delta \rm{M}_\star}{\Delta t} = c_\star \frac{\rm{m}_{gas}}{t_{dyn}}
\end{equation}
where $\Delta$M$_\star$ is the mass of the stars formed in $\Delta t$, 
the time between star formation events (0.8\,Myr in these simulations), 
m$_{gas}$ is the mass of the gas particle, $t_{dyn}$ is the gas 
particle's dynamical time, $c_\star$ the efficiency of star formation, 
in other words, the fraction of gas that will be converted into stars 
during $t_{dyn}$. Effective star formation rates are determined by the 
combination and interplay of $c_\star$ and feedback, and so degeneracies 
do exist between feedback energy and the value of $c_\star$. In this 
study, $c_\star$ is ultimately the free parameter that sets the balance 
of the baryon cycle of cooling gas, star formation, and gas heating. 

Two types of feedback from massive stars are considered, supernovae and 
early stellar radiation feedback.  Supernova feedback is implemented 
using the \citet{stinson06} blastwave formalism, depositing 
$10^{51}$~erg into the surrounding ISM at the end of the stellar 
lifetime of stars more massive than 8\,M$_\odot$.  Since stars form from 
dense gas, this energy would be quickly radiated away due to the 
efficient cooling; for this reason, cooling is disabled for particles 
inside the blast region. Metals are ejected from Type~II supernovae 
(SNeII), Type~Ia supernovae (SNeIa), and the stellar winds driven from 
asymptotic giant branch (AGB) stars, and distributed to the nearest gas 
particles using the smoothing kernel \citep{stinson06}, adopting 
literature yields for SNeII \citep{woosley95}, SNeIa \citep{nomoto97}, 
and AGB stars \citep{vg97}. We trace the lifetimes of stars and supernovae and trace elements Fe, O, C, N, Ne, Si, Mg,  depositing those metals formed at each timestep into the neighbouring gas particles.

Metal diffusion is included, allowing 
proximate gas particles to mix their metals, by treating unresolved 
turbulent mixing as a shear-dependent diffusion term \citep{shen10}.  
Metal cooling is calculated based on the diffused metals.
The impact on the structure of resulting simulated ISM was explored 
by \citet{pilkington11}.

Radiation energy feedback from massive stars has been included in our 
simulations. To model the luminosity of stars, a simple fit of the 
mass-luminosity relationship observed in binary star systems by 
\citet{torres10} is used:
\begin{equation}
\frac{L}{L_\odot} = 
\begin{cases}
(\frac{M}{M_\odot})^{4},  & M < 10 M_\odot\, \\
100(\frac{M}{M_\odot})^{2},   & M > 10 M_\odot\, \\
\end{cases}
\end{equation}
Typically, this relationship leads to 2$\times$10$^{50}$~erg of energy 
being released from the high mass stars per M$_\odot$ of the entire 
stellar population over the $\sim$4.5\,Myr between a star's formation and 
the commencement of SNeII in the region. These photons do not couple 
efficiently with the surrounding ISM \citep{freyer06}.  To mimic this 
highly inefficient energy coupling, we inject 10\% of the energy as 
thermal energy in the surrounding gas, and cooling is \emph{not} turned 
off. Such thermal energy injection is highly inefficient at the spatial 
and temporal resolution of cosmological simulations 
\citep{katz92,kay02}, as the characteristic cooling timescales in the 
star forming regions are lower than the dynamical time. Over $90$\% is 
typically radiated away within a single dynamical time, meaning that our 
effective efficiency of coupling radiation energy feedback to the ISM is 
$\sim$1\%.

As with other galaxy formation simulations in the literature, galaxy 
properties are not precisely the same at different resolutions when the 
same parameters are used \cite[e.g.][]{scannapieco12}. We aim to retain 
the same baryon cycle at the different resolutions as this drives the 
simulated galaxy properties. To achieve this we adjusted our free 
parameter c$_\star$, in order to ensure that each galaxy matches the 
M$_*$-M$_{halo}$ relation at $z$$=$$0$ \citep{moster10}.

\subsection{Turbulent Metal Diffusion}
\label{diffusion}

Galactic inflows and outflows should be turbulent and thus mixing is 
essential for intergalactic medium (IGM) studies. SPH does not 
implicitly include diffusion of scalar quantities such as metals, 
resulting in physically incorrect consequences 
\citep[e.g.][]{wadsley08,shen10,pilkington12b}. By contrast, Eulerian 
grid codes mix due to the necessary advection estimates 
\citep[e.g.][]{few12}. To generate similar non-radiative galaxy cluster 
entropy profiles with SPH as with high resolution grid codes, 
\citet{wadsley08} include a diffusion coefficient, $D$$=$$C_{{\rm 
Diff}}\,\Delta\,v\ h_{\rm SPH}$ based on the pairwise velocity, 
$\Delta\,v$, at the resolution scale, $h_{\rm SPH}$.  A coefficient 
value of order $0.05-0.1$ is expected from turbulence theory (depending 
on the effective measurement scale, $h$).  A conservative choice of 
$C_{{\rm Diff}}$$=$$0.05$ was sufficient to match the cluster comparison 
\citep{wadsley08}, solving a major discrepancy between SPH and grid 
codes \citep{frenk99}. A similar scheme has been applied to supernova 
remnants \citep{greif09}.

\textsc{gasoline} now uses a more robust mixing estimator 
\citep{shen10}, similar to that proposed by \citet{smagorinsky63} for 
the atmospheric boundary layer. Rather than simply using velocity 
differences \citep{wadsley08,greif09}, the diffusion coefficient is now 
calculated according to a turbulent mixing model:
\begin{eqnarray}
\frac{dA}{dt}|_{\rm Diff} & = & \nabla (D \nabla A), \nonumber \\
D & = & C_{{\rm Diff}}\,|S_{ij}|\,h^2,
\end{eqnarray}
where $A$ is any scalar.  The diffusion expression for a scalar $A_p$ on 
particle $p$ is computed as:
\begin{eqnarray}
\tilde{S}_{ij}|_p &=& \frac{1}{\rho_p}\sum_q m_q (v_j|_q-v_j|_p)
\nabla_{p,i} W_{pq}, \nonumber \\ S_{ij}|_p &=& \frac{1}{2}
(\tilde{S}_{ij}|_p+\tilde{S}_{ji}|_p) - \delta_{ij} \frac{1}{3}\ {\rm
Trace}\ \tilde{S}|_p, \nonumber \\ D_p &=& C_{{\rm Diff}}\ |S_{ij}|_p|\ h_p^2, \\
\frac{dA_p}{dt}|_{\rm Diff} &=& -\sum_q m_q
\frac{(D_p+D_q)(A_p-A_q)(\mathbf{r}_{pq}\cdot\nabla_p
W_{pq})}{\frac{1}{2}(\rho_p+\rho_q)\,\mathbf{r}_{pq}^2}, \nonumber
\end{eqnarray}
where the sums are over 32 SPH neighbours, $q$, $\delta_{ij}$ is the 
Kronecker delta, $W$ is the SPH kernel function, $\rho_q$ is the 
density, $\mathbf{r}_{pq}$ is the vector separation between particles, 
$v_i|_q$ is the particle velocity component in direction $i$, $\nabla_p$ 
is the gradient operator for particle p, and $\nabla_{p,i}$ is the {\it 
i}th component of the resultant vector. As the coefficient depends on 
the velocity shear, it better models the mixing in shearing flows. Where 
there is no shearing motion between two phases of fluids, such as for 
compressive or purely rotating flows, no turbulent diffusion is added.

For our fiducial models, we follow the (conservative) choice of 
\cite{shen10}, $C_{{\rm Diff}}$$=$$0.05$. We also run two simulations with 
significantly lower diffusion, $C_{{\rm Diff}}$$=$$0.01$, to test whether 
this diffusion term is driving our results.


.

\section{Results}
\label{results}
Table\,\ref{tab:data} lists the properties of the simulated galaxies, 
including the mass of gas particles, 
star formation efficiency (c$_*$),  metal diffusion coefficient 
(C$_{\rm Diff}$),  total halo mass (M$_{\rm halo}$), virial radius (R$_{\rm vir}$),   stellar mass 
(M$_*$), gas mass (M$_{\rm gas}$), and  R-band magnitude ($M_R$)

We also show in Table\,\ref{tab:data} the final ($z$$=$$0$) baryon mass fraction, 
as compared with the universal value of 0.153, and the fraction of 
baryons that are ever accreted to within the virial radius. In all 
galaxies with M$_{\rm vir}$$>$$8$$\times$$10^{10}$, we do not find any 
indication of prevention of baryons being accreted.  Baryons are 
subsequently ejected from the simulated galaxies, with 50\% ejected from 
SG2 (M$_{\rm vir}$$=$$8.3$$\times$$10^{10}$\,M$_\odot$), decreasing to 18\% of 
gas ejected for L*G2 (M$_{\rm vir}$$=$$7.8$$\times$$10^{11}$\,M$_\odot$). In 
the two lowest mass galaxies, dG3 and SG1, gas is actually prevented 
from being {\rm dm}ed across R$_{\rm vir}$.  In fact over half the baryons 
are never detected within the virial radius of the lowest mass galaxy 
(dG3: M$_{\rm vir}$$=$$2.2$$\times$$10^{10}$\,M$_\odot$). This is due to a 
combination of the UV background radiation and the effect of large scale 
outflows affecting accreting gas.

We also show the baryon fraction that accretes to the star forming 
region: note that we compare to the universal value expected within 
$R_{\rm vir}$. As is also shown in Table~\ref{tab:data}, in the lowest mass 
case, only 14\% of the baryons expected within the virial radius 
actually accrete to the star forming region. This rises to 78\% in SG3 
(M$_{\rm vir}$$=$$1.8$$\times$$10^{11}$\,M$_\odot$) and 80\% in SG4 (M$_{\rm 
vir}$$=$$3.2$$\times$$10^{11}$\,M$_\odot$), before dropping to 69\% in L*G2, possibly due 
to the shock heating of gas in this most massive galaxy, which is above 
the critical mass for shock heating \citep{dekel06}.  Such shock heating occurs in simulations such as ours \citep{keres05,brooks09}.

The inflows and outflows are all self consistently modelled in our simulations, so the outflows we measure in this section  have been affected by the inflows, and the inflows indeed affected by the outflows. We present here net inflows and outflows, as well as quantifying the recycling through galactic fountains. The direction of outflows versus inflows in disc galaxies, and its relation to angular momentum, was studied for these simulations in \cite{brook11}, which showed the preference for outflows to be perpendicular to the disc while inflows are preferentially in the plane of the disc. 

\begin{figure}
\hspace{-.8cm} \includegraphics[height=.68\textheight]{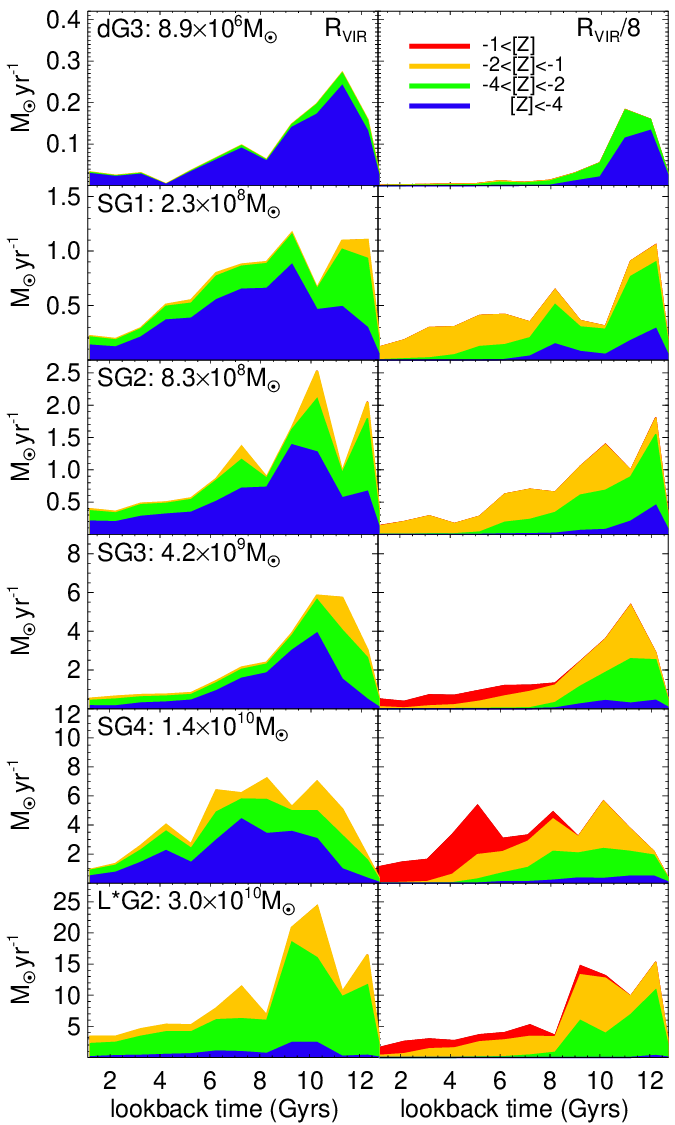}
\caption{Inflow rates (M$_\odot$/yr) of gas entering R$_{\rm vir}$ (left 
panels) and R$_{\rm vir}/8$ (right panels) for the first time, as a function of time. The 
rates are broken into 4 metallicity bins $[Z]$$<$$-4$ (blue), $-4$$<$$[Z]$$<$$-2$ 
(green), $-2$$<$$[Z]$$<$$-1$ (yellow), and $-1$$<$$[Z]$ (red). The stellar mass at z=0 for each simulated galaxy is shown in the left panels. }
\label{inflow}
\end{figure}

\subsection{Inflows}
\label{inflows}

In Figure~\ref{inflow}, we show the inflow rates (M$_\odot$/yr) of gas 
entering R$_{\rm vir}$ (left panels) and R$_{\rm vir}/8$ (right panels) for the first time, 
as a function of time. The rates are broken into four metallicity bins 
$[Z]$$<$$-4$ (blue), $-4$$<$$[Z]$$<$$-2$ (green), $-2$$<$$[Z]$$<$$-1$ (yellow), and $-1$$<$$[Z]$ 
(red).  R$_{\rm vir}$/8 is used to delineate the star forming region of 
the galaxy. Here we have used [Z]$\equiv$log Z/Z$_\odot$

\begin{figure}
\hspace{-.4cm} \includegraphics[height=.43\textheight]{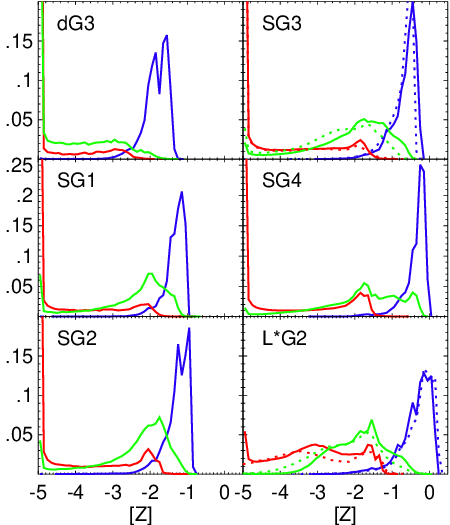}
\caption{Metallicity distribution functions (MDFs) of stars at z=0 
(blue line), along with the MDFs of the gas from which those stars were 
born, measured when the gas first accreted to the virial radius, R$_{\rm 
vir}$ of the galaxy (red line), and when it first reaches the 
star forming region, R$_{\rm vir}$/8 (green line). Gas with 
[Z]$<$$-$5 is shown in the plots at [Z]$=$$-$5. 
Significant amounts of primordial gas is accreted to (R$_{\rm vir}$) in 
the low mass galaxies. Dotted lines in the top right and bottom right panels 
show the minor effect of a five-fold decrease in the metal diffusion 
coefficient (see Section~\ref{lowdiff}).}
\label{MDF}
\end{figure}

In the lowest mass case\footnote{The low mass simulations have higher resolution, and thus our results are not due to an inability to resolve sub-structure.} (dG3), primordial gas enters R$_{\rm vir}$ at 
all times, and most gas entering the star forming region is also 
primordial. As we move to higher masses, more gas is pre-enriched prior 
to entering the virial radius of the central galaxy. This pre-enrichment 
results from the outflows from sub-halos that have collapsed and formed 
stars prior to a lookback time of 12.6~Gyr ($z$$=$$5.5$), enriching the IGM 
from which the central galaxy draws its gas. This finding is consistent 
with what was found in \cite{shen10}, who examined a simulation of 
similar mass as L*G2, with our new study showing that this 
pre-enrichment is also important in lower mass galaxies, although 
progressively less so as we move to lower masses. The fraction of 
primordial gas accreted to the virial radius between lookback times of 
12.5~Gyr ($z$$\sim$$5.5$) and 10~Gyr ($z$$\sim$$2$) goes from 95\% for a 
simulation with M$_*$=8.9$\times$10$^{6}$\,M$_\odot$, to 8\% for 
M$_*$=3.0$\times$10$^{10}$\,M$_\odot$, as shown in Table~\ref{tab:data}. 
It is only in the very lowest mass simulation that subhalo progenitors 
are too low in mass to form stars, prevented by the ionising UV 
background radiation. In all halos above total mass 
$2.2$$\times$10$^{10}$\,M$_\odot$, local collapsing over-densities 
outside the central galaxy result in subhalo progenitors that form stars 
and enrich the IGM during the hierarchical build-up of the galaxies.

Inflow rates peak between $\sim$10$-$12 Gyr ago (3$<$$z$$<$2), in all 
cases except SG4, which has a late merger which brings in significant 
amounts of gas $\sim$8~Gyr ago ($z$$\sim$1).

The most important aspect of Figure~\ref{inflow} is possibly the 
differences between the inflows through R$_{\rm vir}$ (left panels) and 
through R$_{\rm vir}$/8 (right panels).  Firstly, not all gas accreted 
to R$_{\rm vir}$ cools to the star forming region (R$_{\rm vir}$/8), 
with values ranging from $\sim$40\% in the lowest mass halo, to 
$\sim$70\% in the three highest mass cases. The most striking feature 
though is the difference between the metallicity of gas as it accretes 
to R$_{\rm vir}$ for the first time compared to when it accretes to the 
star forming region. The metallicity of the gas has been enhanced during 
its trajectory through the galactic halo, as it mixes with gas that has been enriched in proto-galaxies and by galactic fountains.

Figure~\ref{MDF} summarises the metallicity of the baryons as they 
accrete and subsequently form stars.  The metallicity distribution 
functions (MDFs) of the stars at $z$=0 (blue line)\footnote{Note the 
similarity between the $z$=0 MDF of SG3 and that of 11mChab from 
\citep{pilkington12b}; the two simulations differ only in their adopted 
star formation efficiencies $c_\ast$.} is shown along with the MDFs of 
the gas from which those stars were born, measured when the gas first 
accreted to the virial radius of the galaxy (red line), and when it 
first reaches the star forming region (green line). Gas with 
[Z]$<$$-$5 is taken as primordial and is shown in the plots at 
[Z]$<$$-$5.  Significant amounts of primordial gas is accreted 
to R$_{\rm vir}$ in the low mass galaxies.  Pre-enrichment in 
non-central proto-galaxies is significant in L$_*$ galaxies. The metals 
contained in the gas particles that were not previously within 
non-central proto-galaxies has come from mixing due to metal diffusion 
from metals of the gas pre-enriched in proto-galaxies.

\begin{figure}
\hspace{-.7cm} \includegraphics[height=.375\textheight]{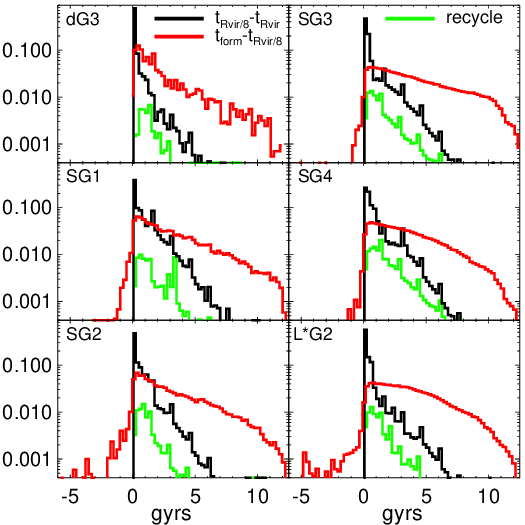}
\caption{Time taken between crossing the virial radius, t$_{\rm{Rvir}}$
and accretion to the star forming region, t$_{\rm{Rvir/8}}$ (i.e.  t$_{\rm{Rvir/8}}$ - t$_{\rm{Rvir}}$, black 
line) for all gas which subsequently forms stars at t$_{\rm{form}}$. We also show
the time taken between accretion to R$_{\rm vir/8}$ and  t$_{\rm{form}}$ (t$_{\rm{form}}$ - t$_{\rm{Rvir/8}}$ red line). Negative values of 
t$_{\rm{form}}$-t$_{\rm{Rvir/8}}$ indicate accreted stars.  The green line shows the distributions of 
times of the galactic fountain cycle for each galaxy - i.e. the time 
between leaving the star forming region and being re-accreted to the 
star forming region. Each histogram shows the gas mass as a fraction of the total z$=$0 stellar mass.}
\label{timescale}
\end{figure}

\subsection{Gas Timescales within the Corona and Disc}
\label{timescales}

 In Figure~\ref{timescale}, we plot as a black line the time taken between crossing the virial radius 
 and accretion to the star forming region 
(t$_{\rm{Rvir}/8}$-t$_{\rm{Rvir}}$),  for gas which subsequently forms 
stars. For this star forming gas as we also plot the time taken between 
entering the star forming region and the time of star formation 
(t$_{\rm{form}}$-t$_{\rm{Rvir}/8}$, red line). Negative values of 
t$_{\rm{form}}$-t$_{\rm{Rvir}/8}$ indicate accreted stars. It is 
interesting (if not surprising) that the galaxies that are massive 
enough to have accreted stars, all accrete pre-enriched gas even as far 
back as $z$$=$$5.5$, whereas it is only in the lowest mass dwarf galaxy that 
stellar accretion does not occur, meaning that the surrounding IGM is 
not enriched and gas is accreted in a pristine state.

As can be garnered from Figure~\ref{timescale}, gas accretes fairly 
rapidly from the virial radius to the star forming region, generally in 
less than 500\,Myr, but with a significant tail out to 5~Gyr. 
Apparently though, this is long enough for some enrichment to occur. The 
length of time between gas being accreted to the star forming region is 
broad, reflecting the low efficiency of star formation in turning gas 
into stars. This allows gas to enrich significantly and relatively 
uniformly, which results in the narrow metallicity distribution 
functions of the stars, as seen by the blue lines of Figure~\ref{MDF}.

Timescales can be approximated by simple exponential functions, with the 
exponent of the time taken between crossing the virial radius 
and entering the star forming region 
(t$_{\rm{Rvir}/8}$-t$_{\rm{Rvir}}$, black line), ranging from $\sim$1 to 1.5/Gyr, with mean 
$\sim$1.2/Gyr and no clear discernible trend with mass (at least within our 
small sample). The exponent of the time taken between accretion to star 
forming region and forming a star  (red line)  ranges 
from $\sim$2 to 5/Gyr, with mean $\sim$4.3/Gyr and some suggestion that lower 
mass galaxies have a shallower slope, but again a larger statistical 
sample will be required.

\subsection{Accreted Stars}
As mentioned above,  negative values of 
t$_{\rm{form}}$-t$_{\rm{Rvir}/8}$ in figure~\ref{timescale} indicate accreted stars.  Few stars accrete directly from sub-halos, and these generally have low metallicity and make up the low metallicity tail of the stellar MDF, and most end in the low mass stellar halo. Gas with low metallicity that falls to the disc generally does not form stars straight away (as shown in Figure 3), so it has higher metallicity (generally) by the time it forms stars. This is why there are there few stars with log(Z/Zsun) below $-$3, in line with observations, i.e. this is the reason that we not have a g-dwarf problem in these simulations. The simulations do make a small number of stars with log(Z/Zsun)$<$-3.

 All simulated galaxies more massive than dG3 clearly show stellar accretion in Figure 3. The timescale over which accretion occurs depends on the particular merger history of each galaxy. This is reflected in that fact the SG2 and L$^*$G2 have, by design, the same merger histories and show similarities in the timescales of star formation for their accreted stars, i.e. the negative values of the red lines of Figure~3. 

\subsection{Recycling of Gas through the Corona}
\label{recycle}

\begin{figure}
\hspace{-.7cm} \includegraphics[height=.4\textheight]{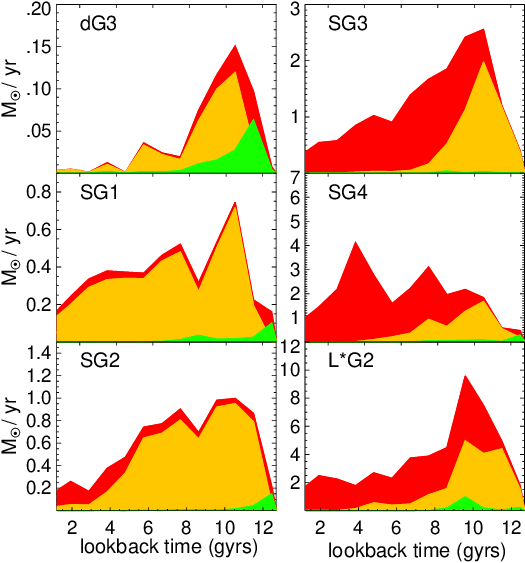}
\caption{Outflow rates (M$_\odot$/yr) of gas through R$_{\rm vir}/8$, as 
a function of time. The rates are broken up into three metallicity bins: 
$-4$$<$$[Z]$$<$$-2$ (green), $-2$$<$$[Z]$$<$$-1$ (yellow), and $-1$$<$$[Z]$ (red).}
\label{outflow}
\end{figure}

Significant numbers of stars are formed from gas that, subsequent to 
accretion to the star forming region, cycles back through the halo in a 
`galactic fountain' prior to eventually forming stars.  The fraction of 
star forming gas involved in galactic fountains, for our simulation 
suite, is shown in column 13 of Table~\ref{tab:data}. Although a larger 
statistical sample may be necessary to fully investigate the dependence 
on mass, our study indicates a balance where, in the lowest mass galaxy, 
only 22\% of stars form from gas that was involved in a fountain, 
presumably because outflows tend to be blown further and do not 
generally re-accrete. At the other end of the mass spectrum (L*G2), 
gravity inhibits large-scale fountains of gas reaching beyond R$_{\rm 
vir}/8$, with 18\% of stars formed from such gas. It is in intermediate 
masses where the galactic fountain is most prevalent, with up to 39\% of 
stars forming from gas cycled through a galactic fountain in a galaxy 
with halo (stellar) mass of $8.3$$\times$10$^{10}$\,M$_\odot$ 
($4.5$$\times$10$^{8}$\,M$_\odot$). We note  that these fractions 
 are sensitive to the chosen radius. 

We plot, as a green line in Figure~\ref{timescale}, the distributions of 
times of the galactic fountain cycle for each galaxy - i.e. the time 
between leaving the star forming region and being re-accreted to the 
star forming region. In each case, this has a similar distribution to 
the timescale for accreting from the virial radius to the star forming 
region. 

The recycle timescales can be loosely fit by exponentials which have 
exponents within 15\% of those of the time taken from the virial radius 
to the star forming region, ranging from $\sim$$-$1 to $-$1.7/Gyr, with a 
mean of $\sim$$-$1.3/Gyr.
Somewhat surprisingly, the timescales of the galactic fountains do not show any strong trend with galaxy mass. 

We are not able to resolve the interaction between clouds of fountain gas and the hot corona with the same detail as e.g. \cite{marinacci10}, where lower metallicity coronal gas becomes entrained with the fountain gas, and thus the inflowing fountain gas has lower metallicity than the outflows. We have simply traced the  gas particles which flow out of,  and back into, the star forming region, whilst any gas swept up in this fountain would simply have been included within our accounting of inflowing gas. We do note, however, that the re-accreted gas particles experience a small loss of metals due to metal diffusion during this cycle, indicating that they are mixing with the lower metallicity halo gas  through the fountain. 


 \begin{figure}
\hspace{-.5cm} \includegraphics[height=.4\textheight]{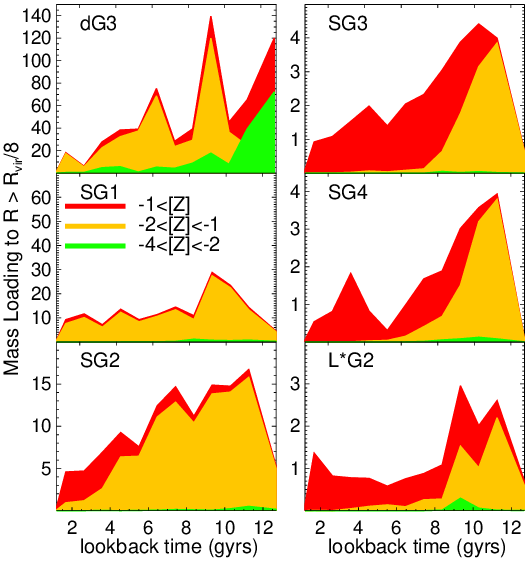}
\caption{Mass loadings: outflow rates (M$_\odot$/yr) of gas through R$_{\rm vir}/8$, divided by the star formation rates, as 
a function of time. These ``mass loadings"  are broken up into three metallicity bins: 
$-4$$<$$[Z]$$<$$-2$ (green), $-2$$<$$[Z]$$<$$-1$ (yellow), and $-1$$<$$[Z]$ (red).}
\label{outflow}
\end{figure}

\subsection{Outflows}
\label{outflows}

To what degree are metals removed from the star forming region during a 
galaxy's evolution? In Figure~\ref{outflow}, we show the outflow rates 
(M$_\odot$/yr) of gas from the star forming region - i.e., outflow rates 
through R$_{\rm vir}/8$, as a function of time, corresponding to the 
amount of metals blown into regions that allow pollution of the 
infalling gas. The rates are broken up into three metallicity bins: 
$[Z]$$<$$-4$ (blue), $-4$$<$$[Z]$$<$$-2$ (green), and $-2$$<$$[Z]$$<$$-1$ (yellow). Outflows 
are significantly enriched with metals, and are enriching the 
inter-stellar, cirum-galactic, and inter-galactic media.

\subsection{Mass Loading}
We can also plot the outflows in terms of the mass loading factor, which measures outflow rates compared to the star formation rates. Firstly, we note again that outflow rates {\it are} sensitive to  the radius through which one measures outflows: in \cite{brook12a}, we showed that the distribution of distances that outflowing gas travels from the star forming region drops off exponentially. 
We are interested in the outflows from the star forming region, as this is what helps determines the availability of gas and metals for subsequent star formation. We therefore have measured the mass loading by determining outflows rates through R$_{\rm vir}$/8. 

As can be seen, the mass loading factors for the lowest mass galaxies are high, with gas outflowing from the star forming region at rates that are factors of 100 greater than the star formation rates at high redshift.  Mass loadings are galaxy mass dependant and redshift dependant, with the highest mass loadings occurring in the lowest mass galaxies at high redshift. The highest  mass simulation, the L$^*$ galaxy, has had gas outflows from the star forming region at rates similar to the star formation rate since $z$$\sim$1. 

\subsection{Effective Yields}
The effective yield measures how a galaxyÕs metallicity deviates from what would be expected for a closed box model of galaxy formation, i.e. a galaxy with the same gas mass fraction that had evolved without inflow or outflow of gas. A closed box galaxy evolution obeys a simple relationship between  gas metallicity and the gas mass fraction.
 Thus, effective yields place constrains on the baryon cycle, which must have a combination of inflows and outflows of gas and metals that result in  matching observed values of effective yields. The effective yield is defined as:
 
$$y_{eff}\equiv \frac{Z_{\rm gas}}{\rm{In} (1/f_{\rm gas})}, $$

\begin{figure}
\hspace{-.5cm} \includegraphics[height=.19\textheight]{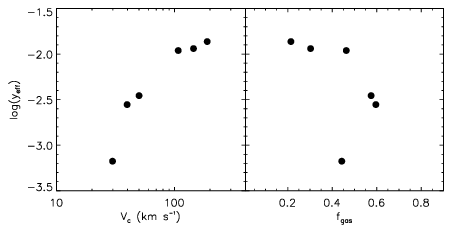}
\caption{We plot the effective yields, $y_{\rm eff},$ as a function of rotation velocity (V$_c$, left panel) and  gas fraction (f$_{\rm gas}$, right panel). }
\label{ey}
\end{figure}

 
 Clearly, our simulations are far from closed box models as our earlier analysis has shown, and we measure here how such deviations  are reflected in the effective yields of our simulations. Figure~\ref{ey} shows the effective yields as a function of rotation velocity (V$_c$) and as a function of gas fraction (f$_{\rm gas}$). A flattening of the relation for galaxies with $V_c$$>100$\,kms$^{-1}$ is also seen in observations \citep{pilyugin04}, although we note an offset in  effective yields between the observations and our simulations, which may be due to metallicity calibrations. The trend of effective yields with gas fraction is also  similar between the simulations and observations.




\subsection{Effect of Metal Diffusion}
\label{lowdiff}

In Figure~\ref{MDF}, we overplot as dotted lines the metallicity 
distribution functions (MDFs) for the two low diffusion runs showing 
stars at $z$$=$$0$ (blue dotted lines), along with the MDFs of the gas from 
which those stars were born, measured when the gas first accreted to the 
virial radius (R$_{\rm vir}$, red dotted lines), and when it first 
reaches the star forming region, (R$_{\rm vir}$/8, green dotted 
lines). Gas with [Z]$<$$-$5 is shown in the plots at 
[Z]$=$$-$5. The runs with a low value for the metal diffusion 
coefficient ($C_{{\rm Diff}}$$=$$0.01$) are consistent with the 
results presented previously, where we used the fiducial $C_{{\rm 
Diff}}$$=$$0.05$, indicating that the choice of diffusion coefficient is 
not driving our results.  We note that the complete elimination of
diffusion leads to unrealistic MDF dispersions, skewness, and 
kurtosis \citep{pilkington12b}.

\section{Discussion}
\label{summary}

The flow of gas and metals into, out of, and around galaxies comprises 
the baryon cycle, which ultimately is responsible for setting the 
characteristics observed in galaxies today. Any comprehensive theory of 
galaxy formation must aim for a constrained baryon cycle at its heart. 
Chemical evolution models attempt to do so by employing 
spatially-resolved abundance patterns, while dynamical 
models attempt to match galaxy scaling relations, amongst other things. In this study, we have explored the cycle of 
baryons and metals within a suite of simulated galaxies that match a 
range of galaxy scaling relations, having been tuned to match the 
M*$-$M$_{halo}$ relation \citep{brook12b}. Our simulations add energy only locally in star forming regions, which drives outflows along the path of least resistance \cite[see e.g.][]{brook11}. No scaling of outflows with galaxy mass or star formation rate is input by hand, nor is the direction of the outflows. Inflowing gas can penetrate and provide gas to the star forming regions, even while outflows are occurring \cite[e.g.][]{brooks09}. 
We have traced the infall of 
gas from $z$$=$$5.5$ to the present day, to the star forming regions of 
simulated galaxies that span a range in stellar mass between 
M$_*$=8.9$\times$10$^{6}$\,M$_\odot$ and 
M$_*$=3.0$\times$10$^{10}$\,M$_\odot$. We have traced the cycle of gas 
through the galactic fountain for each of the galaxies, and also shown 
the rate at which metals are removed from the star forming region.

We find that the universal baryon fraction accretes to the virial radius 
in all galaxies with halo (stellar) mass $\geq$ $8.3$$\times$10$^{10}$ 
($4.5$$\times$10$^{8}$)\,M$_{\odot}$. Lower mass galaxies have gas 
prevented from accreting by the background UV radiation field, and 
perhaps due to outflows, with only $\sim$$46\%$ of the universal baryon 
fraction accreting to the virial radius of a galaxy of virial mass 
$2.2$$\times$10$^{10}$\,M$_\odot$. Significant baryonic outflows occur in each 
simulation, with more outflows in low mass systems resulting in the baryon content at $z$$=$$0$ monotonically increasing 
with mass, ranging from 15\% to 82\% of the universal value in our 
fiducial runs. Not all the baryon baryonic content within R$_{\rm}$ ever reach the star forming 
region (R$_{\rm vir}$/8): only 14\% of the universal value in the lowest 
mass case, rising to $\sim$$80$\% for galaxies with virial mass $\sim$$3\times$10$^{11}$\,M$_\odot$, while the amount of 
gas reaching the star forming region in the most massive galaxy (M$_{\rm 
vir}$$=$$7.6\times$10$^{11}$\,M$_\odot$) is lower,  69\%, we speculate that this is due to  shock heating at the virial radius (see \citealt{keres05,brooks09}).

None of the simulated galaxies have a significant population of low 
metallicity stars, with relatively narrow metallicity distribution 
functions at all masses (n.b., see also \citealt{pilkington12b} and 
\citealt{calura12}). Enrichment of infalling gas occurs in three stages: 
(i) pre-enrichment in progenitor sub-halos, (ii) enrichment after 
accretion to the virial radius but subsequent to accretion to the star 
forming region, and (iii) enrichment in the star forming region.

\begin{enumerate}
\item {\it Pre-enrichment in progenitor sub-halos:} In the very 
lowest mass halo  there is no star formation within progenitor sub-halos, 
and hence only in the lowest mass halo is accreted gas not enriched by 
sub-halos polluting the IGM prior to redshift $z$=5.5. The amount of gas 
pre-enriched in this manner is a strong function of galaxy mass. Accretion of pre-enriched gas goes hand in hand with the existence of an accreted halo component. 
\item {\it Enrichment after accretion to the virial radius but 
subsequent to accretion to the star forming region:} In all simulations, 
metals are ejected from the star forming region, largely to the 
surrounding hot halo. As fresh gas accretes through the virial radius it 
is mixed with the metals of this hot enriched halo. Further, the effect 
of metals is to  decrease cooling times, which has an effect in selecting more 
metal enriched gas to preferentially accrete from the hot halo to the 
star forming region. The result is that gas accreted to the star forming 
region is significantly more enriched than gas accreted through the 
virial radius. The distribution of timescales for gas crossing the 
virial radius and passing to the star forming region is reasonably 
approximated by an exponential with exponent $\sim$$-$1.2/Gyr.
\item {\it Enrichment in the star forming region:} Low star formation 
efficiency in disc galaxies results in a broad distribution in the time 
that baryons spend between first being accreted to the star forming 
region, and then subsequently forming stars. This gas may remain in the 
star forming region, or cycle through the hot halo and re-accrete to the 
star forming region to form stars (see \citealt{brook12a}). The result 
is that gas accreted to the star forming region is significantly more 
enriched than gas crossing the virial radius. The distribution of 
timescales between gas passing to the star forming region and finally 
forming a star is reasonably approximated by an exponential with 
exponent $\sim$$-$4.3/Gyr.
 \item {\it Maintaining low metallicity :}  Low star formation 
efficiency in low mass galaxies galaxies, and a low star to gas ratio means that enrichment  of the ISM proceeds slowly, 
allowing low mass simulated galaxies to maintain low metallicity despite outflows being sub-solar metallicity \cite[see][]{dalcanton07}.

\end{enumerate}

A significant number of stars in each galaxy form from gas that 
undergoes at least one {\it cycle through the galactic fountain}, 
ranging from $\sim$$20$$-$$40\%$, with intermediate mass halos (M$_{\rm 
vir}$$\sim$$8$$\times$$10^{10}$\,M$_\odot$) having the most such stars: gas in 
lower mass galaxies has an increasing  tendency to be expelled without being re-accreted, while at the more massive en
 the larger potential well in higher mass galaxies means that gas 
is not blown as far into the hot halo. The distribution of timescales 
for which gas remains within a galactic fountain is similar to that of 
the time taken from crossing the virial radius to arriving at the star 
forming region. As we showed in 
\cite{brook12a}, the distance that fountain gas reaches from the centre 
of the galaxy drops off exponentially. Hence, significantly more star forming gas 
could be involved in smaller-scale galactic fountains, particularly 
perpendicular to the disc, in the manner of e.g. \citet[][]{marinacci11}.

\section*{Acknowledgments}
CBB is supported by the MICINN (Spain) through the grant AYA2009-12792.
 GS and AVM acknowledge support from SFB 881 (subproject A1) of the DFG. CBB acknowledges Max- Planck-Institut f\"{u}r Astronomie for its hospitality and financial support through the Sonderforschungsbereich SFB 881 ÒThe Milky Way SystemÓ (subproject A1) of the German Research Foundation (DFG).
 We acknowledge the computational support 
provided by the UK's National Cosmology Supercomputer (COSMOS), as well as the theo cluster of the Max-Planck-Institut f\"{u}r Astronomie at the Rechenzentrum in Garching.  We thank the DEISA consortium, co-funded through EU FP6 project RI-031513 and the FP7 project RI-222919, for support within the DEISA Extreme Computing Initiative, the UKÕs National Cosmology Super-computer (COSMOS), and the University of Central LancashireÕs High Performance Computing Facility.

\bibliographystyle{mn2e}
\bibliography{brook}

\end{document}